\documentclass[conference,a4paper]{IEEEtran}
\usepackage[nolist]{acronym}
\usepackage{tikz}
\usetikzlibrary{decorations.pathreplacing}
\usetikzlibrary{arrows.meta}
\usetikzlibrary{patterns}
\usetikzlibrary{shapes,arrows}
\usetikzlibrary{calc}
\usepackage{listings}
\usepackage{mathtools}
\usepackage[hyphens]{url}

\usepackage{cite}

\ifCLASSINFOpdf
\else
\fi
\begin{document}

\begin{acronym}
\acro{abi}[ABI]{Application Binary Interface}
\acro{aslr}[ASLR]{Address Space Layout Randomization}
\acro{bmbf}[BMBF]{Federal Ministry of Education and Research}
\acro{bsa}[BSA]{Base Save Area}
\acro{dsp}[DSP]{Digital Signal Processing}
\acro{cisc}[CISC]{Complex Instruction Set Computer}
\acro{esa}[ESA]{Extra Save Area}
\acro{gpu}[GPU]{Graphics Processing Unit}
\acro{jop}[JOP]{Jump-Oriented-Programming}
\acro{lr}[LR]{Link Register}
\acro{mmu}[MMU]{Memory Management Unit}
\acro{pc}[PC]{Program Counter}
\acro{risc}[RISC]{Reduced Instruction Set Computer}
\acro{rop}[ROP]{Return-Oriented-Programming}
\acro{sp}[SP]{Stack Pointer}
\acro{tie}[TIE]{Tensilica Instruction Extension}
\acro{xea2}[XEA2]{Xtensa Exception Architecture 2}
\acro{xea3}[XEA3]{Xtensa Exception Architecture 3}
\end{acronym}
%
\title{Challenges of \acl{rop} on the Xtensa Hardware Architecture}

\author{\IEEEauthorblockN{Kai Lehniger\IEEEauthorrefmark{1}, Marcin J. Aftowicz\IEEEauthorrefmark{1}, Peter Langend\"orfer\IEEEauthorrefmark{1}\IEEEauthorrefmark{2}, Zoya Dyka\IEEEauthorrefmark{1}}
\IEEEauthorblockA{\IEEEauthorrefmark{1}IHP - Leibniz-Institut f\"ur innovative Mikroelektronik\\
Frankfurt (Oder), Germany\\
Email: \{lehniger, aftowicz, langendoerfer, dyka\}@ihp-microelectronics.com}
\IEEEauthorblockA{\IEEEauthorrefmark{2}Brandenburgische Technische Universit\"at Cottbus-Senftenberg\\
Cottbus, Germany\\
Email: peter.langendoerfer@b-tu.de}}


\IEEEspecialpapernotice{{\normalfont\scriptsize(Author's version accepted for DSD-2020; the final publication is available at https://ieeexplore.ieee.org/abstract/document/9217805)}}

\maketitle

\begin{abstract}
This paper shows how the Xtensa architecture can be attacked with \acf{rop}. The presented techniques include possibilities for both supported \acfp{abi}. Especially for the windowed \ac{abi} a powerful mechanism is presented that not only allows to jump to gadgets but also to manipulate registers without relying on specific gadgets. This paper purely focuses on how the properties of the architecture itself can be exploited to chain gadgets and not on specific attacks or a gadget catalog.
\end{abstract}


%
\IEEEpeerreviewmaketitle

\section{Introduction}
Cadence Tensilica IP includes Tensilica Xtensa extensible processors \cite{CadenceDesignSystems.2020}, a broad range of high-performance application-specific DSPs and a comprehensive development toolchain supporting fast and easy hardware and software tradeoff ana\-ly\-sis for digital signal processing and embedded control. 
By the end of 2012, already 2 billion IP cores were shipped by licensees \cite{sembyotic.com.20200618T12:32:58.000Z}, around 800 million cores per year and with an increasing rate. 
\par
With \acf{rop}, a technique generalized from return-into-libc attacks by \cite{Shacham.2007}, an attacker makes use of already existing code snippets that end with a return instruction. These code snippets are called gadgets and can be used as jump targets instead of the shellcode. By chaining multiple gadgets together it is possible to execute arbitrary code. It has been shown that ROP chaining is turing complete for large enough programs \cite{Shacham.2007,Homescu.2012}.\par
While finding gadgets can be considered similar to other architectures, chaining these gadgets highly depends on the \acf{abi}. Making it more difficult, the Xtensa processor can be configured with two different \acp{abi}. This paper will highlight the important parts of both \acp{abi} that make \ac{rop} possible. While for both \acp{abi} \ac{rop} is possible, there are differences on how to construct the payload and the kind of gadgets that can be used.\par
The rest of this paper is structured as follows: section \ref{sec:related} gives an overview of the related work, section \ref{sec:rop} explains the concept of \ac{rop} for the x86 architecture, section \ref{sec:xtensa} roughly explains some details of the Xtensa needed to understand the attack while sections \ref{sec:call0} and \ref{sec:windowed} focuses on the attacks for the two different \acp{abi}, and section \ref{sec:conclusion} concludes this paper.

\section{Related Work}\label{sec:related}
Stack buffer overflows have been a known vulnerability for several decades now\cite{AlephOne.1996}. They allow an attacker to manipulate program state or inject shellcode. One of the more frequently manipulated values is the return address of a function. It could for example be used to jump into injected shellcode. While first buffer overflow attacks were shown for the x86 architecture, McDonald showed that the same attack was possible on the SPARC architecture \cite{McDonald.1999}. SPARC contains a sliding register window, which made the attack need one more return before it could work. The first return is used to load the manipulated return address from the stack into the \%i7 register. The second return will then jump to that manipulated address.\par
With the broad introduction of W$\oplus$X protection for processors, shellcode injection was no longer possible. Return-into-libc attacks were a first form of code reuse attacks that targeted libc functions instead of shellcode. This approach was practical since libc was loaded into almost every C written application. Later, \ac{rop} was introduced as a generalization of this technique \cite{Shacham.2007} using gadgets instead of libc functions as jump target.\par
While the attack was introduced for the x86 architecture first, it was adapted for \ac{risc} architectures too \cite{ErikBuchanan.2008,Kornau.2009,Checkoway.2010,Davi.2010}. \cite{ErikBuchanan.2008} describes the attack for the SPARC architecture by manipulating local and input registers and provides a gadget catalogue. In \cite{Kornau.2009,Checkoway.2010,Davi.2010} the attack was shown for the ARM architecture by using different instructions than return, i.e. \texttt{pop pc}, but with the same result. In \cite{klog.14.02.2020} a frame pointer that was saved on the stack was manipulated to shift the top of the stack back into the payload. The same approach is used in this paper to chain gadgets and optimize the size of the payload for the windowed \ac{abi}.\par
In 2018 \ac{rop} attacks for the Xtensa were already demonstrated \cite{VanRooyen.2018} but without making use of the full potential of the windowed \ac{abi} and concluded that this \ac{abi} makes it harder to find gadgets. They recommended \ac{jop} attacks instead. In contrast, this paper shows more potential for the windowed \ac{abi}, at least for register manipulation, making regular \ac{rop} attacks possible.
\section{\acl{rop} - the classics}\label{sec:rop}
The return address of a function is used to remember where the function was called from and to hand the control flow back to that position once the function is done. Because there is an arbitrary number of functions and functions can also be recursive the space for return addresses cannot be calculated statically. Therefore the return address is put on the stack.\par
On the x86 architecture the \texttt{call} instruction, that is used to call subroutines, automatically pushes the return address to the stack, while the \texttt{ret} instruction, that is used to return from subroutines, pops the return address from the stack. An attacker can, by abusing a stack buffer overflow vulnerability, overwrite the return address on the stack with a new value of his choice. In case of \ac{rop} this new value is the address of a gadget. Because this gadget itself also ends with a \texttt{ret} instruction it can jump again to a new location that can be controlled by the attacker. This is working because the \texttt{ret} instruction does two important things: first it takes an address from the top of the stack, and second it increments the stack pointer. Without the second part the stack pointer would not change and another \texttt{ret} instruction would use the same return address, if no other instructions would change the stack pointer. Chaining different gadgets without this mechanism would be a more difficult task.\par
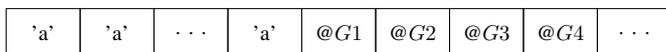
\begin{figure}[b!]
\centering
 \resizebox{\linewidth}{!}{\begin{tikzpicture}[scale=1]
\def\w{1.2cm}
\def\h{.8cm}
\def\x{0}
\def\y{0}
\foreach \t [count=\i] in {'a','a',. . .,'a',@$G1$,@$G2$,@$G3$,@$G4$,. . .} 
  \draw (\x+\i*\w,\y)node[draw,minimum height=\h,minimum width=\w]{\t};
\end{tikzpicture}}
\caption{Simple payload for an \ac{rop} attack on the x86 architecture}
\label{fig:x86_payload}
\end{figure}

Because of these properties the assembly of the attackers payload is pretty straight forward. \figurename~\ref{fig:x86_payload} shows a simple example. First, the buffer must be filled with random bytes up to the point of the first return address on the stack. After  that the gadget addresses can simply be put one after another since the \texttt{ret} instruction of the end of every gadget will point the stack pointer to the next injected gadget address.\par
Of course the payload becomes more complex when other information is integrated into it, e.g. function arguments or values for \texttt{pop} instructions.
Due to the similarities of the architectures it is worth to mention the attack of a SPARC machine described in \cite{ErikBuchanan.2008} due to the similarities of the architectures. They exploit the window sliding mechanism by overwriting the local and input register values stored on the stack. Because stack pointer and return address are stored in the input registers, both values can be controlled. This paper will describe a similar technique for the Xtensa architecture in more detail.
\section{Xtensa}\label{sec:xtensa}
This section describes the parts of the Xtensa architecture that are relevant for \ac{rop}.\par
Xtensa has a configurable and extensible architecture, designed to be optimized for a specific application. For example it is possible to extend the instruction set with the \ac{tie} language. In this paper however, only a configuration with the basic instruction set is covered. The basic instruction set consists of 24-bit instructions with 16-bit (.n narrowed) versions for the most common instructions to save code size. Custom instructions could be even longer, up to 128 bits. The different instruction sizes can be intermixed with each other. This allows to find gadgets on other positions than function epilogues, which is especially important for the windowed \ac{abi}.\par
Most important for this paper, the configuration allows to select between two software \acp{abi}, called \textbf{call0} and \textbf{windowed}. This selection fundamentally changes the way how gadgets must be chained.\par
Xtensa is a 32 bit architecture with a word width of 4 bytes. All entries in figures of this paper that represent registers or values in the stack are the size of a word.
Xtensa provides a set of 16 general registers a0 to a15. a0 is used to store the return address and a1 holds the function's \ac{sp}.\par
All the necessary information regarding the Xtensa architecture addressed in this paper can also be found in greater detail in 
 \cite{CadenceDesignSystems.2019}.
\section{call0}\label{sec:call0}
\subsection{Architecture}
In the call0 \ac{abi}, if a function calls a subroutine, the calling conventions demand that a0 must be saved on the stack and restored before the return. This is done during the function's prologue and epilogue. The frame of a typical non-leaf function is shown in Listing~\ref{lst:function_frame}. In the prologue, at the beginning of the function, stack space is reserved for the function by decrementing the \ac{sp}. Then the return address is stored on the stack. The epilogue reverses this operation. Since there are no push or pop instructions in the instruction set, load and store instructions are used. The position of the return address on the stack is referenced with the \ac{sp} and an offset, in this example 8. This position is not defined and can differ from compiler to compiler and function to function. 
\par
\begin{figure}[b!]
\begin{lstlisting}[label=lst:function_frame,caption=call0 function frame,captionpos=b]
// prologue
addi	a1, a1, -16 // reserve stack space
s32i.n	a0, a1, 8   // store return address
// function body
...
// epilogue
l32i.n	a0, a1, 8   // load return address
addi	a1, a1, 16  // free stack frame
ret.n
\end{lstlisting}
\end{figure}
Functions itself are called either with the \texttt{call0} or \texttt{callx0} instruction. 
Return is either done by the \texttt{ret} or \texttt{ret.n} instruction.

\subsection{Gadget Chaining Mechanism}
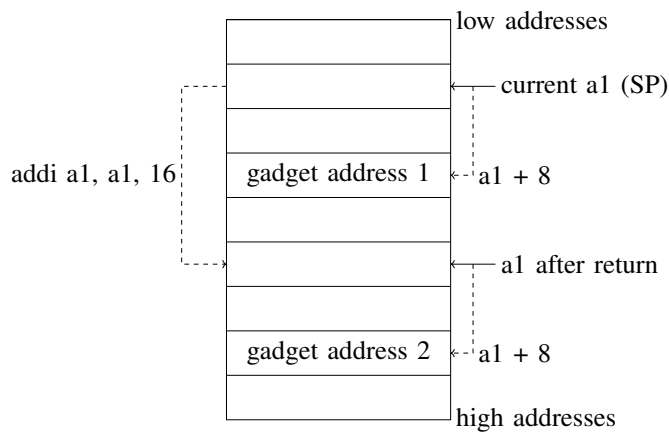
\begin{figure}[b!]
\centering
 \resizebox{\linewidth}{!}{\begin{tikzpicture}[scale=1]
\tikzstyle{every node}=[font=\LARGE]
\def\s{9}
\def\w{5}
\draw (0,0)--(0,-\s)--(\w,-\s)node[right]{high addresses}--(\w,0)node[right]{low addresses}-- cycle;
\pgfmathparse{\s-1}
\foreach \x in {0,...,\pgfmathresult} 
  \draw (0,-\x)--(\w,-\x)(\w/2,-\x-.5);
\draw[<-] (\w, -1.5)--(\w+1,-1.5)node[right]{current a1 (\ac{sp})};
\draw[<-] (\w, -5.5)--(\w+1,-5.5)node[right]{a1 after return};
\draw[->,dashed] (\w+.5,-1.5)--(\w+.5,-3.5)node[right]{a1 + 8}--(\w,-3.5);
\draw[->,dashed] (\w+.5,-5.5)--(\w+.5,-7.5)node[right]{a1 + 8}--(\w,-7.5);
\draw[->,dashed] (0, -1.5)--(-1, -1.5)--(-1,-5.5)node[midway,left]{addi a1, a1, 16}--(0,-5.5);
\draw (\w/2,-3.5)node{gadget address 1};
\draw (\w/2,-7.5)node{gadget address 2};
\end{tikzpicture}}
\caption{Epilogue execution}
\label{fig:tracking}
\end{figure}
The mechanism of a pop from the stack, or a similar operation, explained in section \ref{sec:rop}, must be replicated. The first task is to load the return address from the stack. The second task is to change the position of the stack pointer. Since both tasks are not done by the \texttt{ret} instruction directly, other instructions need to do that in every gadget.\par
For the call0 \ac{abi} function epilogues can be used as gadgets. Since they load the return address back into a register and also move the \ac{sp}, they do all what is needed to chain gadgets. When building the payload for a \ac{rop} attack it is important to know where the addresses of the gadgets must be placed within the payload. The \texttt{l32i.n} instruction (see Listing~\ref{lst:function_frame}) is used to restore the return address to a0. It uses the \ac{sp} (a1) plus an offset to address the position on the stack where the return address was stored. Therefore the current value of a1 (or at least its change over time) and the offset must be known to the attacker. a1 is mainly changed in function prologues and epilogues. Since the prologue decrements a1, which is the wrong direction if we want to make use of the payload of the buffer overflow, it should be avoided in gadgets. In the epilogue, an \texttt{addi} instruction is used to increment a1. All that needs to be done is to keep track of this increments in all gadgets to determine the position for the next gadget address in the payload. \figurename~\ref{fig:tracking} illustrates how this can be done. It shows a stack that was already overwritten by a buffer overflow. To make it as easy as possible the two gadgets that are executed consist only of the epilogue shown in Listing~\ref{lst:function_frame}. So first a0 is restored by using the position of a1 with an offset of 8. This is where the address of the first gadgets must be placed. After that a1 is incremented by 16 and the gadget returns to the next gadget. This one, again, uses the same offset. Since a1 changed by 16, we have to place the next gadget address 4 words apart the previous one in the payload.

\section{windowed}\label{sec:windowed}
\subsection{Sliding window Mechanism}
With the windowed \ac{abi} a function also has access to the general registers a0 to a15. However, in this case these registers are just a register window of a larger physical register bank with a size of either 32 or 64. 
How much the window slides depends on the call instruction that is used. In contrast to the call0 \ac{abi} there are 6 different call instructions that can be used: \texttt{call}$n$ and \texttt{callx}$n$ where $n$ denotes how many registers the window slides. 
For example a \texttt{call4} instruction will cause the window to slide down $4$ registers. The return is done by an \texttt{retw} (windowed return) or \texttt{retw.n}, which slides the window back.\par
However, the sliding itself is not done by the \texttt{call}(\texttt{x})$n$ instruction, but by the \texttt{entry} instruction. The \texttt{entry} instruction is required to be the first instruction in every function and replaces the instructions form the call0 prologue. This is necessary because sliding the window down by i.e. 4 would cause a0 to a3 (or more) to get out of the current window. Since the stack pointer rests in a1, it would no longer be possible to decrement it and reserve space on the stack. First the old value of a1 is saved, then the window slides and after that the decremented value is written back into new a1. For the return address this problem is solved in the way that \texttt{call}(\texttt{x})$n$ simply puts the return address into a$n$, which will be a0 after the window slide. However, the caller cannot reserve stack space for the callee. 
With the windowed \ac{abi} there is no need for an epilogue in every non-leaf function anymore. In the call0 \ac{abi} the standard epilogue only restores a0 and a1 which is done automatically in the windowed \ac{abi} when window slides back up.\par
Both, \texttt{entry} and \texttt{retw}, must know which \texttt{call}(\texttt{x})$n$ instruction was used to slide the window. This information is stored in the two most significant bits of the return address by the call instruction. During the return the \texttt{retw} will substitute these two bits by the current \ac{pc}'s two most significant bits. Therefore it is not possible to return to a different 1GB segment in memory with \texttt{retw}.\par

\subsection{Register spilling}
Due to the limited number of physical registers, window sliding cannot go on indefinitely. When new registers are needed and no free physical registers are available, a register overflow exception occurs. This causes the call of an overflow handler which saves the affected register's current values on the stack. Therefore every window register has its predefined position in the stack where it can be saved to. The area of the stack frame, reserved for this purpose, is called Window Save Area. This area is comprised of two parts, the \ac{bsa} for the registers a0 to a3 with a fixed size of 16 bytes, and the \ac{esa} for the registers a4 to a11. This area can have a size of 0, 16, or 32 bytes.\par
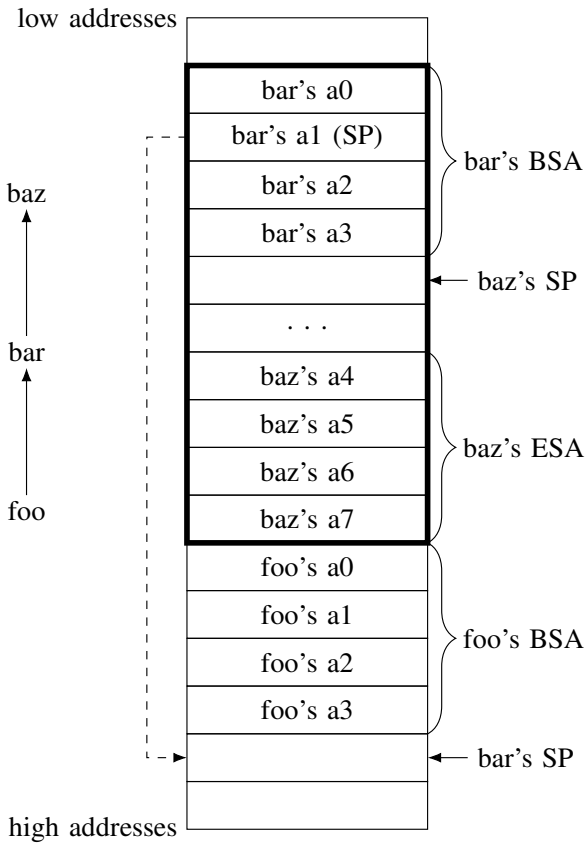
\begin{figure}[t!]
\centering
 \resizebox{.9\linewidth}{!}{\begin{tikzpicture}[scale=1]
\def\w{3cm}
\def\elems{10}
\def\hs{.6};

\foreach \x [count=\i] in {,bar's a0,bar's a1 (SP), bar's a2, bar's a3,,. . .,baz's a4, baz's a5, baz's a6, baz's a7, foo's a0, foo's a1, foo's a2, foo's a3,,} 
  \draw (0,-\i*\hs)node[draw,minimum height=\hs cm,minimum width=\w]{\x};
  
\draw[line width=2] (-\w/2,-\hs *1.5) rectangle (\w/2,-\hs * 11.5);
  
\def\p{16};
\def\pp{6};
\def\high{17};
\draw[{Latex}-](\w/2,-\hs*\p)--(\w/2+.5cm,-\hs*\p)node[right]{bar's SP};
\draw[{Latex}-](\w/2,-\hs*\pp)--(\w/2+.5cm,-\hs*\pp)node[right]{baz's SP};
\draw(-\w/2,-\hs/2)node[left]{low addresses};
\draw(-\w/2,\high*-\hs-\hs/2)node[left]{high addresses};

\draw[-{Latex},dashed](-\w/2,-\hs*3)--(-\w/2-.5cm,-\hs*3)--(-\w/2-.5cm,-\hs*\p)--(-\w/2,-\hs*\p);

\def\o{1.5}
\def\l{3.5}
\draw[-{Latex}] (-\l,-5-\o)node{foo}(-\l,-4.8-\o)--(-\l,-3.2-\o);
\draw[-{Latex}] (-\l,-3-\o)node{bar}(-\l,-2.8-\o)--(-\l,-1.2-\o);
\draw(-\l,-1-\o)node{baz};

\draw [decorate,decoration={brace,amplitude=10pt}](\w/2,-1.5*\hs) -- (\w/2,-5.5*\hs) node [black,midway,xshift=1.2cm] {bar's BSA};
\draw [decorate,decoration={brace,amplitude=10pt}](\w/2,-7.5*\hs) -- (\w/2,-11.5*\hs) node [black,midway,xshift=1.2cm] {baz's ESA};
\draw [decorate,decoration={brace,amplitude=10pt}](\w/2,-11.5*\hs) -- (\w/2,-15.5*\hs) node [black,midway,xshift=1.2cm] {foo's BSA};
\end{tikzpicture}}
\caption{Register positions on stack for \texttt{call8}}
\label{fig:saved_registers}
\end{figure}

\figurename~\ref{fig:saved_registers} shows the reserved positions of some registers for the functions foo, bar, and baz. It assumes that the function foo called bar, which in turn called baz. It shows the stack frame for the function baz (highlighted by the black box). All functions were called with the \texttt{call8} instruction. Therefore the size of the Extra Save Area is 16 bytes, to contain the registers a4 to a7.\par
First it must be noted that the \ac{bsa} for a function is not located on the functions own stack frame. This is due to the fact that a1 is used as the stack pointer. Because all stack interaction is done through the stack pointer it is not possible to save the first 4 registers in a function's own stack frame. That's why the \ac{bsa} of foo is located in bar's stack frame and bar's \ac{bsa} is located in baz's stack frame. They are stored with a negative offset to the stack pointer. This way it is also possible to backtrace the stack, because the saved stack pointer in a stack frame points to the next stack frame on the stack.\par
The \ac{esa} is located on a function's own stack frame on its highest addresses. However, because the size of a stack frame is variable for each function (indicated by the . . .) and these registers are stored at the beginning of the stack frame, a function's stack pointer cannot be used to address these positions with a fixed offset. But they have a fixed offset to the previous function's stack pointer. This is why bar's stack pointer is used to restore baz's registers a4 to a7. This would be the same for the variable window variant and a \texttt{call12}, with the exception that the registers a4 to a11 would be saved at the beginning of the stack frame. It should be noted that in this case the registers a4 to a7 have a different offset to the stack pointer because they got displaced by the registers a8 to a11. This fact is important when building the payload for the attack.

\subsection{Gadget Chaining Mechanism}
Since the function epilogue is mostly replaced by the sliding window mechanism, the method described earlier for the call0 \ac{abi} is no longer working. However, the mechanics of the \texttt{retw} instruction itself can be exploited. When registers have been spilled to the stack and have to be reloaded on return, the underflow exception handler can be used to load manipulated register values back into the current register window. With that both essential registers for return address (a0) and stack pointer (a1) can be manipulated by the attacker. There are 3 different handlers, one for each possible window slide. They restore 4, 8, or 12 registers. Because the decision which handler is called is based on the two most significant bits in the return address, it is possible for the attacker to select how many registers should be loaded from the stack back to the register bank. Therefore the values of a0 to a11 can be controlled directly just by using the \texttt{retw} instruction without the need of any actual gadget. The remaining registers a12 to a15 can be controlled too, because of the slide mechanism. It just takes more then one gadget to control them. One gadget that loads the registers a8 to a11 and the second which just slides the register window by 4, or similar constellations.\par
The underflow exception handlers restore the registers by making use of the fact that the stack can be backtraced to restore the registers a4 to a11. Therefore this property must be reconstructed in the payload that is injected by an attacker. For this, a linked list structure for the payload is introduced. It is a fitting data structure because the saved stack pointer that points to the next stack frame behaves similar to a next pointer of a list element. \figurename~\ref{fig:window_payload} shows the beginning of such a list with 3 list elements. The structure of a list element is defined by the stack structure that was introduced in the previous subsection, because all the offsets to the stack pointers must be kept identical. Each list element mimics a \ac{bsa} on the stack. A potential \ac{esa} is indicated by the dotted parts. This is increasing the size of a list element. An underflow exception that is raised within \texttt{retw}, will use this injected \ac{bsa} (and \ac{esa}) to restore the registers and with that load the next gadget address (marked as @$Gi$ for gadget $i$). Each register has a gadget as subscript to indicate what gadget is being executed when the register value is loaded. The difference of the placement of gadget addresses and its register values that can be observed is caused by the fact that it takes two returns to execute a gadget. The first return will just load the gadget address into the register a0 while the second return will jump to that gadget address. At the same time the underflow exception of the second return will also load the next gadget address and other registers that the first gadget can use. This is why the register values for the first gadget must be placed in the second list element. The registers a4 to a11 are placed in a different list element from the first 4 registers because they are part of the \ac{esa}, which is part of a different stack frame from the corresponding \ac{bsa}. While a0 to a3 are always restored, the other registers are only loaded if the WindowUnderflow8 or WindowUnderflow12 handlers are used. Also, the positions of the register a4 to a7 differs between these two handlers and must be taken into account.\par
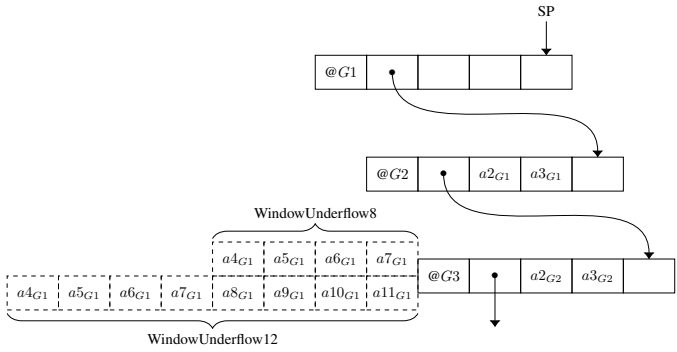
\begin{figure}[t!]
\centering
 \resizebox{\linewidth}{!}{\begin{tikzpicture}[scale=1]
\def\w{1.2cm}
\def\h{.8cm}
\def\x{0}
\def\y{0}
\foreach \t [count=\i] in {@$G1$,$\bullet$,,,} 
  \draw (\x+\i*\w,\y)node[draw,minimum height=\h,minimum width=\w]{\t};
\def\x{1*\w}
\def\y{-3*\h}
\foreach \t [count=\i] in {@$G2$,$\bullet$,$a2_{G1}$,$a3_{G1}$,} 
  \draw (\x+\i*\w,\y)node[draw,minimum height=\h,minimum width=\w]{\t};
\def\x{2*\w}
\def\y{-6*\h}
\foreach \t [count=\i] in {@$G3$,$\bullet$,$a2_{G2}$,$a3_{G2}$,} 
  \draw (\x+\i*\w,\y)node[draw,minimum height=\h,minimum width=\w]{\t};
\def\x{-2*\w}
\def\y{-5.5*\h}
\foreach \t [count=\i] in {$a4_{G1}$,$a5_{G1}$,$a6_{G1}$,$a7_{G1}$} 
  \draw (\x+\i*\w,\y)node[dashed,draw,minimum height=\h,minimum width=\w]{\t};
\def\x{-6*\w}
\def\y{-6.5*\h}
\foreach \t [count=\i] in {$a4_{G1}$,$a5_{G1}$,$a6_{G1}$,$a7_{G1}$,$a8_{G1}$,$a9_{G1}$,$a10_{G1}$,$a11_{G1}$} 
  \draw (\x+\i*\w,\y)node[dashed,draw,minimum height=\h,minimum width=\w]{\t};
  
\def\arrow{-{Latex[length=1.5mm,width=2.4mm]}};
\draw[\arrow](5*\w,1.5*\h)node[above]{SP} to (5*\w,.5*\h);
\draw[\arrow](2*\w,0) to [out = 270,in = 90, looseness = 1] (6*\w,-2.5*\h);
\draw[\arrow](3*\w,-3*\h) to [out = 270,in = 90, looseness = 1] (7*\w,-5.5*\h);
\draw[\arrow](4*\w,-6*\h) to (4*\w,-7.5*\h);

\draw [decorate,decoration={brace,amplitude=10pt}](-1.5*\w,-4.9*\h) -- (2.5*\w,-4.9*\h) node [black,midway,yshift=.6cm] {WindowUnderflow8};
\draw [decorate,decoration={brace,amplitude=10pt,mirror}](-5.5*\w,-7.1*\h) -- (2.5*\w,-7.1*\h) node [black,midway,yshift=-.6cm] {WindowUnderflow12};
\end{tikzpicture}}
\caption{Payload as linked list}
\label{fig:window_payload}
\end{figure}
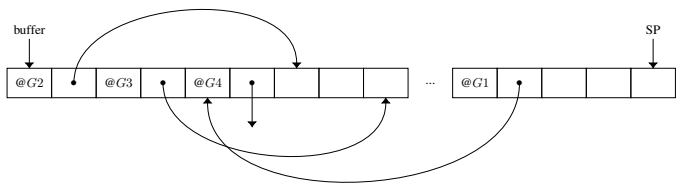
\begin{figure}[b!]
\centering
 \resizebox{\linewidth}{!}{\begin{tikzpicture}[scale=1]
\def\w{1.2cm}
\def\h{.8cm}
\def\x{0}
\def\y{0}
\foreach \t [count=\i] in {@$G2$,$\bullet$,@$G3$,$\bullet$,@$G4$,$\bullet$,,,} 
  \draw (\x+\i*\w,\y)node[draw,minimum height=\h,minimum width=\w]{\t};
\def\x{10*\w}
\def\y{0}
\foreach \t [count=\i] in {@$G1$,$\bullet$,,,} 
  \draw (\x+\i*\w,\y)node[draw,minimum height=\h,minimum width=\w]{\t};
\draw (10*\w,0)node[]{...};  
\def\arrow{-{Latex[length=1.5mm,width=2.4mm]}};
\draw[\arrow](1*\w,1.5*\h)node[above]{buffer} to (1*\w,.5*\h);
\draw[\arrow](15*\w,1.5*\h)node[above]{SP} to (15*\w,.5*\h);

\draw[\arrow](12*\w,0) to [out = 270,in = 270, looseness = 1] (5*\w,-.5*\h);
\draw[\arrow](2*\w,0) to [out = 90,in = 90, looseness = 1] (7*\w,.5*\h);
\draw[\arrow](4*\w,0) to [out = 270,in = 270, looseness = 1] (9*\w,-.5*\h);

\draw[\arrow](6*\w,0) to (6*\w,-1.5*\h);

\end{tikzpicture}}
\caption{Linked list as packed payload}
\label{fig:packed_payload}
\end{figure}
Packing the payload is also a bit different from other architectures. The most important part is the placement of the first list element because this must align with the \ac{sp} of a function which returns with an underflow exception. The rest of the list elements can be placed at will. They can even overlap each other as long as the overlapping values are either the same or do not matter for the affected gadgets. But the biggest advantage is that the beginning of the payload, which normally contains junk bytes to fill the buffer, can be used to store list elements. This way attacks with just a handful of gadgets can fit almost entirely inside the buffer, resulting in very small payloads, assuming a reasonable sized buffer. A minimalistic example that does not care about any registers, but a0 and a1, is shown in \figurename~\ref{fig:packed_payload}. While the first gadget address and the first next pointer must be placed further down the stack, the rest of the chaining is done in the original buffer. Gadget addresses and next pointers simply alternate, with the next pointer being incremented by two words each time.

\subsection{Differences to the SPARC architecture}
The described attack and the attack on the SPARC described in \cite{ErikBuchanan.2008} share a lot of similarities. In both cases the restore mechanism of the sliding register window is attacked. But differences occur in details. The biggest advantage of the Xtensa architecture is its flexibility, which unfortunatly can be used by the attacker. While on the SPARC the register window always shifts by 16 registers, allowing 8 register values to be handed over to the next gadget, Xtensa allows shifts by 4, 8, or 12 registers. Since the attacker is able to control the window shift it is easier for him to pass values between gadgets and keep them in the register window for a longer time when using a window shift of 4. A drawback from the attacker's point of view is that the assembly of the attack becomes more complicated because the window shift is an additional variable that needs to be taken into account. However, \cite{ErikBuchanan.2008} solved this problem of not being able to passing values between many gadgets by storing all variables in memory. Of course this technique is also applicable for Xtensa.
\section{Conclusion}\label{sec:conclusion}
In this paper we showed how the Xtensa architecture can be attacked with \ac{rop} no matter what \ac{abi} is used in the configuration. We proposed a method for the windowed \ac{abi} that allows arbitrary register manipulation combined with an efficient packing of the payload. This makes regular \ac{rop} attacks possible for the windowed \ac{abi}. This potential can be used additionally to the instructions of the gadget and needs more research to be fully explored, for example if and how an integration in existing \ac{rop} compilers is possible.

\section*{Acknowledgment}
This work was supported by the \ac{bmbf} under research grant number 01IS18065E.



\bibliographystyle{IEEEtran}
\bibliography{IEEEabrv,bib}

\begin{thebibliography}{10}
\providecommand{\url}[1]{#1}
\csname url@samestyle\endcsname
\providecommand{\newblock}{\relax}
\providecommand{\bibinfo}[2]{#2}
\providecommand{\BIBentrySTDinterwordspacing}{\spaceskip=0pt\relax}
\providecommand{\BIBentryALTinterwordstretchfactor}{4}
\providecommand{\BIBentryALTinterwordspacing}{\spaceskip=\fontdimen2\font plus
\BIBentryALTinterwordstretchfactor\fontdimen3\font minus
  \fontdimen4\font\relax}
\providecommand{\BIBforeignlanguage}[2]{{%
\expandafter\ifx\csname l@#1\endcsname\relax
\typeout{** WARNING: IEEEtran.bst: No hyphenation pattern has been}%
\typeout{** loaded for the language `#1'. Using the pattern for}%
\typeout{** the default language instead.}%
\else
\language=\csname l@#1\endcsname
\fi
#2}}
\providecommand{\BIBdecl}{\relax}
\BIBdecl

\bibitem{CadenceDesignSystems.2020}
{Codence Design Systems, Inc.}, ``{Tensilica Customizable Processors},''
  https://ip.cadence.com/ipportfolio/tensilica-ip/xtensa-customizable (April
  2020).

\bibitem{sembyotic.com.20200618T12:32:58.000Z}
\BIBentryALTinterwordspacing
S.~sembyotic.com, ``Tensilica licensees ship two billion ip cores; license
  revenue now larger than any other dsp licensing company | cadence ip,''
  2020-06-18T12:32:58.000Z. [Online]. Available:
  \url{\url{https://ip.cadence.com/news/400/330/Tensilica-Licensees-Ship-Two-Billion-IP-Cores-License-Revenue-Now-Larger-than-Any-Other-DSP-Licensing-Company}}
\BIBentrySTDinterwordspacing

\bibitem{Shacham.2007}
H.~Shacham, ``The geometry of innocent flesh on the bone: Return-into-libc
  without function calls (on the x86),'' in \emph{Proceedings of the 14th ACM
  conference on Computer and communications security}, 2007, pp. 552--561.

\bibitem{Homescu.2012}
A.~Homescu, M.~Stewart, P.~Larsen, S.~Brunthaler, and M.~Franz, ``Microgadgets:
  size does matter in turing-complete return-oriented programming,'' in
  \emph{Proceedings of the 6th USENIX conference on Offensive Technologies},
  2012, p.~7.

\bibitem{AlephOne.1996}
{Aleph One}, ``Smashing the stack for fun and profit,'' \emph{Phrack Magazine,
  49(14)}, 1996.

\bibitem{McDonald.1999}
\BIBentryALTinterwordspacing
J.~McDonald, ``Defeating solaris/sparc non-executable stack protection,''
  Bugtraq mailing list, 1999. [Online]. Available:
  \url{https://seclists.org/bugtraq/1999/Mar/4}
\BIBentrySTDinterwordspacing

\bibitem{ErikBuchanan.2008}
{Erik Buchanan}, {Ryan Roemer}, {Hovav Shacham}, and {Stefan Savage}, ``When
  good instructions go bad: Generalizing return-oriented programming to risc,''
  in \emph{In Proceedings of the 15th ACM Conference on Computer and
  Communications Security (CCS}, 2008.

\bibitem{Kornau.2009}
T.~Kornau, ``Return oriented programming for the arm architecture,'' 2009.

\bibitem{Checkoway.2010}
S.~Checkoway, L.~Davi, A.~Dmitrienko, A.-R. Sadeghi, H.~Shacham, and
  M.~Winandy, ``Return-oriented programming without returns,'' in
  \emph{Proceedings of the 17th ACM conference on Computer and communications
  security}, 2010, pp. 559--572.

\bibitem{Davi.2010}
L.~Davi, A.~Dmitrienko, A.-R. Sadeghi, and M.~Winandy, ``Return-oriented
  programming without returns on arm.''

\bibitem{klog.14.02.2020}
\BIBentryALTinterwordspacing
klog, ``The frame pointer overwrite,'' 14.02.2020. [Online]. Available:
  \url{http://www.phrack.org/issues/55/8.html}
\BIBentrySTDinterwordspacing

\bibitem{VanRooyen.2018}
C.~{Van Rooyen} and P.~Promeuschel, ``Expanding exploitation beyond x86 and
  arm, into the realm of xtensa,'' in \emph{Nullcon~Goa 2018}, 2018.

\bibitem{CadenceDesignSystems.2019}
I.~{Cadence Design Systems}, ``Xtensa instruction set architecture (isa)
  reference manual,'' 2019.

\end{thebibliography}

\end{document}